\documentclass[10pt, conference, compsocconf]{IEEEtran}
\IEEEoverridecommandlockouts
\usepackage{cite}
\usepackage{amsmath,amssymb,amsfonts}
\usepackage{algorithmic}
\usepackage{graphicx}
\usepackage{textcomp}
\usepackage{xcolor}
\usepackage{multirow}
\usepackage{diagbox}
\usepackage{booktabs}
\usepackage{algorithm}
\usepackage{enumitem}
\usepackage{float}
\usepackage[font=small,skip=0pt]{caption}
\graphicspath{ {./images/} }
\def\BibTeX{{\rm B\kern-.05em{\sc i\kern-.025em b}\kern-.08em
    T\kern-.1667em\lower.7ex\hbox{E}\kern-.125emX}}
\begin{document}

\title{A Hybrid Deep Learning Anomaly Detection Framework for Intrusion Detection}

\author{\IEEEauthorblockN{Rahul Kale}
\IEEEauthorblockA{
%	\textit{dept. name of organization (of Aff.)} \\
\textit{ST Engineering, Singapore}\\
%City, Country \\
RahulVishwanath.Kale@stengg.com}
\and
\IEEEauthorblockN{Zhi Lu}
\IEEEauthorblockA{
%	\textit{dept. name of organization (of Aff.)} \\
\textit{ST Engineering, Singapore}\\
%City, Country \\
lu.zhi@stengg.com}
\and
\IEEEauthorblockN{Kar Wai Fok}
\IEEEauthorblockA{
%	\textit{dept. name of organization (of Aff.)} \\
\textit{ST Engineering, Singapore}\\
%City, Country \\
fok.karwai@stengg.com}
\and
\IEEEauthorblockN{Vrizlynn L. L. Thing}
\IEEEauthorblockA{
	%	\textit{dept. name of organization (of Aff.)} \\
	\textit{ST Engineering, Singapore}\\
%	City, Country \\
	vriz@ieee.org}}

\maketitle

\begin{abstract}
Cyber intrusion attacks that compromise the users' critical and sensitive data are escalating in volume and intensity, especially with the growing connections between our daily life and the Internet. The large volume and high complexity of such intrusion attacks have impeded the effectiveness of most traditional defence techniques. While at the same time, the remarkable performance of the machine learning methods, especially deep learning, in computer vision, had garnered research interests from the cyber security community to further enhance and automate intrusion detections. However, the expensive data labeling and limitation of anomalous data make it challenging to train an intrusion detector in a fully supervised manner. Therefore, intrusion detection based on unsupervised anomaly detection is an important feature too. In this paper, we propose a three-stage deep learning anomaly detection based network intrusion attack detection framework. The framework comprises an integration of unsupervised (K-means clustering), semi-supervised (GANomaly) and supervised learning (CNN) algorithms. We then evaluated and showed the performance of our implemented framework on three benchmark datasets: NSL-KDD, CIC-IDS2018, and TON\_IoT.
\end{abstract}

\begin{IEEEkeywords}
Anomaly Detection, Intrusion Detection, Deep Learning, Unsupervised Learning, Neural Networks
\end{IEEEkeywords}

\section{Introduction}\label{sec:introduction}
Internet based systems are becoming an integral part of the everyday life with the advancements in networking\cite{shafique2020internet} and AI\cite{yang2020federated}. With increased connectivity and technology adoption, there are rising concerns in cyberattacks. Intrusion detection systems serve as an important defense against cyberattacks, to identify network threats and enable appropriate measures to be taken.
Intrusion detection systems based on anomaly detection adopt different approaches such as \cite{kim2019rapp,liu2021hybrid} to handle various types of challenges and datasets \cite{pang2021deep}. Anomaly is essentially a deviation from the rest of the normal data points or network traffic patterns. The similarity among key features of normal samples can be quantified to distill the anomalous ones. Various supervised learning methods \cite{mighan2021novel,  liu2021hybrid} were proposed for this anomaly detection problem. The main challenge is the need to generate a large dataset with annotated labels \cite{chandola2009anomaly}. To alleviate the need of such a large labeled training dataset, unsupervised learning methods are proposed in the literature for anomaly detection \cite{guounsup,fan2020robust,pu2020hybrid, boniol2020sad, li2021deep}.

In this work, we propose an intrusion detection framework using unsupervised anomaly detection that integrates the clustering and deep learning techniques. It consists of three stages: The first stage separates the more obvious anomalies from the normal samples using K-means clustering. These remaining “normal” samples are sent to the second stage, to further identify the anomalies through the adaption of the GANomaly\cite{akcay2018ganomaly}.
GANomaly learns the compressed representation of the input data sample and computes the difference between such data representation and its reconstructed representation. The anomalies are identified if their reconstructed error is larger than the threshold. For the first two stages, we assume that the anomalies do not dominate the dataset statistically, which is the notion of an anomaly\cite{foorthuis2021nature}. We also propose a final-stage Convolutional Neural Network (CNN) classification model. This stage is useful for attack type analysis in the event that annotated attack types are available. We evaluate the performance of our proposed framework on NSL-KDD, CIC-IDS2018, and TON\_IoT datasets.

The rest of the paper is organized as follows. Section II reviews the related works. The details of the proposed framework is presented in Section III. The experimental setup and dataset details are discussed in Section IV. The results are presented and discussed in Section V. Section VI concludes the paper.
\section{Related Works}
Many different approaches for anomaly detection and intrusion detection are found in literature. Mighan and Kahani \cite{mighan2021novel} proposed a combination of stacked auto-encoder network and classifiers such as Support Vector Machine (SVM), Random Forest (RF).
Kim et. al proposed a semi-supervised method called reconstruction along projection pathway (RaPP) for novelty detection in \cite{kim2019rapp}. This method utilized the activation values for hidden space by feeding the auto-encoder its reconstructed input.
A hybrid intrusion detection method was proposed in \cite{liu2021hybrid} which uses a combination of K-means and RF for initial identification of normal and anomaly samples. CNN, Long-Short-Term-Memory (LSTM) were used to identify the attack category of the samples. The evaluation was performed on NSL-KDD and CIC-IDS2017 datasets.
These methods are based on the completely supervised setup and hence require large number of labeled anomalous samples. However, it is not always possible to have such sufficiently large labeled training dataset.
Hence, efforts were focused on to propose unsupervised algorithms for anomaly detection problems.

Guo et al. developed a gated recurrent unit and gaussian mixture VAE scheme in \cite{guounsup} to address the challenge of anomaly detection in IoT systems.
Robust unsupervised anomaly detection methods proposed in \cite{fan2020robust} by Fan et al. combines convolutional AE and Gaussian process regression for feature extraction and anomaly removal from noisy data. Unsupervised anomaly detection method integrating sub-space clustering and one-class SVM is proposed in \cite{pu2020hybrid} by Pu et al. and the performance is verified on the NSL-KDD dataset.
Boniol et al. propose an unsupervised subsequence anomaly detection method which performs domain-agnostic anomaly detection in \cite{boniol2020sad}. The method proposed in \cite{li2021deep} by Li et al. combines clustering and AE for completely unsupervised anomaly detection. Continuous representation learning and reconstruction error calculation through iterations between these two steps provide consistent performance across datasets.

According to the survey \cite{pang2021deep}, deep learning methods improve the recall for the anomaly detection problem.
Therefore, deep learning methods are preferred for the anomaly detection problem in the recent literature \cite{pang2021deep}. To identify the types of the anomalies, at least a partially trained classification network is necessary in the framework. Through a combination of unsupervised, semi-supervised and supervised learning methods in our framework, we aim to address the limitations present in supervised and unsupervised learning methods individually.

\section{Intrusion Detection Framework}
We will first formally formulate the problem for the anomaly detection.
\subsection{Problem Formulation}
Let $\mathcal{D} = \{\mathbf{x}_{i}\}$, $i = 1,\dots,K$ , $\mathbf{x} \in \mathbb{R}^n$ be the entire dataset of the input data samples including the anomalies and normal samples. The objective of the anomaly detection in the second stage of the framework is to generate anomaly scores $h_{i}$ to classify the samples $\mathbf{x}_{i}$ based on a threshold $th$ such that:
\begin{equation*}
	y_{i} =
	\begin{cases}
		0, & \text{if $h_i < th$}\\
		1, & \text{if $h_i \geq th$}\\
	\end{cases}
\end{equation*}
\noindent where $y_i$ are the predicted labels. $y_i = 1$ indicates that the sample $x_i$ is an anomaly and $y_i = 0$ indicates that the sample $x_i$ is normal sample. The details about the score $h_i$ are discussed in Sec.\ref{sec:gano}

We will now describe the three stages from the intrusion detection framework.

\subsection{Stage 1: Pre-processing and Clustering} \label{sec:kmeans}
\subsubsection{Pre-processing}
Prior to performing the clustering step, the input data needs to undergo some pre-processing. The datasets may contain categorical features which are converted to ordinal numeric values. For example, if a categorical feature has 70 distinct values, the numeric equivalents between [1,70] with a step size of 1 will be assigned to the features.
For normalization, in this paper, we adopt min-max normalization method to map the values to the range [0,1].
\subsubsection{Clustering}

The k-means clustering algorithm is utilized here to identify the normal data samples. The main idea here is to select k samples as cluster centers. Then, we calculate the distance between the sample points and the cluster centers. Each point will be assigned a cluster such that the distance between that cluster center and the point will be the least. These steps will be repeated until a prior defined stopping criterion is satisfied. The distance referred in the k-means algorithm is the Euclidean distance.

Once the clustering process finishes, it is critical to select the most probable normal data samples for the training of the adversarial network in the second stage. For this selection process, we consider two selection factors: the cluster size $C$ and the cluster variance. The rationale behind opting for these two factors is the assumption that the normal samples will form larger clusters with smaller cluster variance\cite{li2021deep}. $Th_{sz}$ and $Th_{var}$ represent the selection criterion for the cluster size and variance respectively. First, only the clusters with size greater than $Th_{sz}$ will be the eligible clusters for the normal sample selection. Within each eligible cluster $m$ with cluster size $C_m$, only the $Th_{var} \times C_m$ samples closest to the center of the cluster will be selected and added to the training set of probable normal samples for the next stage of the framework. The complete training set is formed by combining the probable normal samples from each eligible cluster.

\subsection{Stage 2: Generative Adversarial Network} \label{sec:gano}

For the anomaly detection in the second stage, we have adapted the Generative Adversarial Network(GAN) based anomaly detection network proposed by Samet et al. in \cite{akcay2018ganomaly}. GANomaly network mainly consists of three sub-networks. The first sub-network is an auto-encoder style generator network with an encoder and decoder. The purpose of the generator is learning the input data representation through encoder and reconstruction of the input data through the decoder. The second sub-network consists of an encoder network which downscales the reconstructed data from the first sub-network. The encoder in this sub-network is architecturally identical to the encoder in the first sub-network. The discriminator network in the third sub-network has the objective of identifying the input and reconstructed data as real or fake. The objective function of the network is as follows:
\begin{equation*}
	L = w_{adv}L_{adv} + w_{con}L_{con} + w_{enc}L_{enc}
\end{equation*}
\noindent where $w_{adv}$, $w_{con}$ and  $w_{enc}$ are the parameters that adjust the impact of the individual losses $L_{adv}$, $L_{con}$ and  $L_{enc}$, respectively. The adversarial loss $L_{adv}$ is calculated between the feature representation of original and reconstructed data. The contextual loss $L_{con}$ is calculated between the input data and the reconstructed data. The encoder loss $L_{enc}$ is calculated between the compressed vector representations of the input and reconstructed data.

The network in \cite{akcay2018ganomaly} is primarily designed for image inputs and evaluated with image datasets. Therefore, their implementation utilizes 2-D convolutions and 2-D transposed convolutions which are designed to identify the spatial relationships between the adjacent pixels within the image data \cite{sensors}. However, our proposed framework mainly handles the network traffic data which does not have any spatial relationship, as we assume that each captured feature is independent and its position within the input sample vector is not significantly relevant. Hence, unlike the 2-D convolutions in \cite{akcay2018ganomaly}, the 1-D convolutions and 1-D transposed convolutions are utilized in this paper to implement GANomaly in the second stage of the framework.
During the prediction stage, the encoder loss $L_{enc}$ is used for generating the anomaly score of each sample. The equation for calculating $L_{enc}$ is as follows \cite{akcay2018ganomaly}:
\begin{equation*}
	A(\hat{x}) = ||G_E(\hat{x}) - E(G(\hat{x}))||
\end{equation*}
\noindent where for each test sample $\hat{x}$, $A(\hat{x})$ is the anomaly score, $G_E(\hat{x})$ is the compressed vector of the input  and $E(G(\hat{x}))$ is the compressed vector of the reconstruction of the input data. The anomaly scores $S = \{s_i : A({z_i}), {z_i} \in Z\}$ are calculated and feature scaling is applied as shown in the formula below to scale the anomaly scores within the probabilistic range [0,1].
\begin{equation*}
	h_i = \frac{s_i - min(S)}{max(S) - min(S)}
\end{equation*}
\noindent where the score $h_i$ is the anomaly score referred in the problem formulation.

\subsection{Stage 3: Convolutional Neural Network} \label{sec:cnn}
\begin{figure}[htb]
	\centering
	\includegraphics[width=3in]{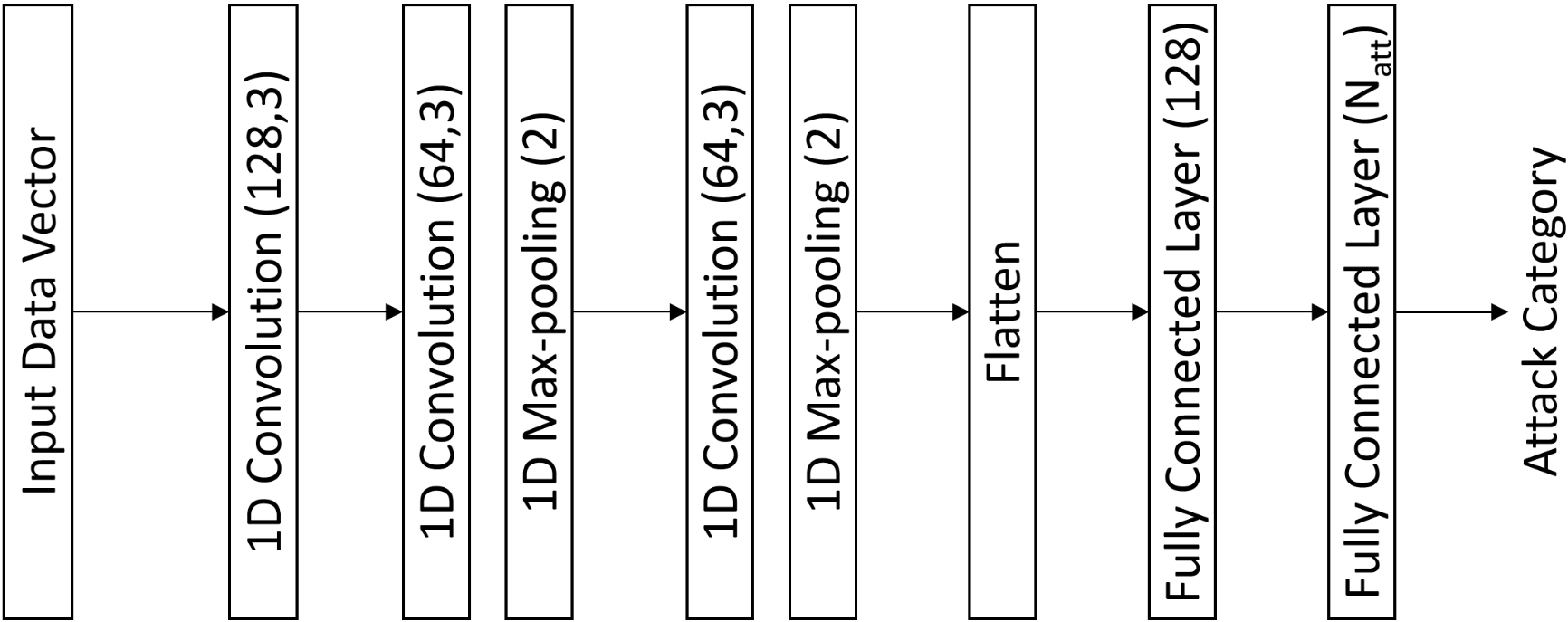}
	\setlength{\belowcaptionskip}{-8pt}
	\caption{CNN Structure}
	\label{fig:block_cnn}
\end{figure}

In the last stage, we designed a CNN model to classify the anomalies in their respective categories. This CNN mainly consist of an input layer, a few convolutional and pooling layers, fully connected layers. The convolution between data and the kernels extracts the features in the convolution layer. Through multiple convolutional layers, many hidden features can be extracted to assist in the classification. To have the more abstract and higher level of features, pooling layers are used on the learned representations. In this paper, we use maxpooling which selects the maximum value from the set of selected feature map values to reduce the overall computations. Rectified Linear Unit (ReLU) will be used as the activation function. In fully connected layer, the activation of each neuron is calculated by comparing the weighted sum of its inputs with a threshold. This neuron is connected to all activations from the previous layer. The general network structure of the CNN used in this paper is shown in Fig. \ref{fig:block_cnn}.
\vspace{-1em}

The input and output shape parameters depend on the dataset used as the number of distinct attack categories for each dataset is different. The numeric values in the bracket for 1D Convolution correspond the number of kernels and the size of each kernel, respectively. The numeric value in the bracket for 1D Max-pooling refers to the pooling size. The $N_{att}$ corresponds to the number of attack categories present in the dataset.

The training of the CNN stage is performed offline. That is, the CNN used in the framework is trained using the labeled samples from the three datasets used in this paper. Though the label information is not utilized for the first two stages of framework, to train the CNN for classification, the label information is necessary. As the input to this third stage are only the anomalies identified from the second stage, training of the CNN was performed using only the anomaly samples of different attack categories i.e. training/test data did not have any normal or benign samples. Adam optimizer and categorical cross-entropy is used for calculating loss function.

The distribution of training sets for different attack categories may not be even which may impact the performance as the samples from the majority classes will dominate the training process. To address the type imbalance problem, ADASYN \cite{adasyn} will be used in this paper. ADASYN is an oversampling algorithm that adaptively generates the samples from the minority class, depending on the probability distribution of the minority class.

\section{Experimental Setup}
The performance of the proposed framework will be evaluated using the three datasets: NSL-KDD\cite{nsl}, CIC-IDS2018\cite{sharafaldin2018toward} and TON\_IoT\cite{moustafa4}. We will briefly describe each dataset.

\begin{itemize}[leftmargin=*]
	\item \textbf{NSL-KDD}\cite{nsl}: NSL-KDD dataset is designed to overcome certain inherent issues from the KDD99 dataset such as redundant, duplicate records; availability of sufficient records in different difficulty levels; and reasonable number of records in both training and testing dataset.
	There are total 43 features within the dataset.
	\item \textbf{CIC-IDS2018}\cite{sharafaldin2018toward}:	
	CIC-IDS2018 dataset includes the system logs of multiple machines and captures of their network traffic, with total 80 features extracted from the captured network traffic. The timestamp feature was dropped from the dataset as ideally the anomaly detection or attack category identification should not depend on the timestamp. There are total 14 attack types.
	\item \textbf{TON\_IOT(Windows10)}\cite{moustafa4}:	
	This dataset consists of network traffic captured from the different IoT and IIoT sensors with Linux and Windows system trace datasets collected from the hosts running the corresponding operating systems. In this paper, we are using the Windows 10 dataset which is collected using the Performance Monitor Tool on Windows 10 systems. The dataset collected activities of desk, process, processor, memory and network activities from the Windows 10 systems. There are total 124 features available within the dataset. The timestamp feature was dropped from the dataset as ideally the anomaly detection or attack category identification should not depend on the timestamp. There are total 7 attack types.
\end{itemize}
The sample distribution details of all three source datasets is given in Table\ref{table:main_samp_distrib}.
\vspace{-1em}
\begin{table}[htbp]
	\centering
	\caption{Sample distribution for source datasets}
	\label{table:main_samp_distrib}
	{\begin{tabular}{|c|c|c|}
			\hline
			Dataset & Normal Samples & Anomalous Samples \\
			\hline
			NSL-KDD & 76432 & 66684 \\
			\hline
			CIC-IDS2018 & 2856035 & 1669364 \\
			\hline
			TON\_IoT (Win10) & 10000 & 11104 \\
			\hline				
			%		\caption{Sample Distribution for NSL-KDD Dataset}
	\end{tabular}}
%	\vspace{-1em}
\end{table}

We will now describe the overall experimental design. Though all the three datasets are labeled, for the first two stages of the framework, the label information will be dropped and will not be used for training the algorithm. Therefore, the combination of the first two stages of the framework will be unsupervised. For the anomaly detection problem (Stage 1 and Stage 2 of the framework), the different attack categories will be treated under single anomaly ``Attack" category.

To rigorously evaluate the effectiveness of the algorithms for anomaly detection, the experiment was designed in the following manner: First, from NSL-KDD dataset, 5 non-overlapping subsets are created randomly. Each subset is referred to as a SampleSet. The normal and anomalous samples in each SampleSet are picked completely randomly from the main NSL-KDD dataset. There is no overlap between either normal or anomalous samples among the SampleSets. The number of normal and anomalous samples in each SampleSet is identical. Anomaly percentage in each SampleSet is maintained around 10. Next, for each SampleSet, 3-fold cross-validation(CV) is performed. The final result is the average of CV results obtained for each SampleSet.

Kindly note that only NSL-KDD dataset is mentioned in the above steps, but these steps are followed for the evaluation of CIC-IDS2018 and TON\_IoT datasets. The size of each SampleSet for the three datasets is given in Table \ref{table:sampset_distrib}.
\vspace{-0.5em}
\begin{table}[htbp]
	\centering
	\caption{SampleSet Distribution for the three datasets. There are total 5 non-overlapping SampleSets for each dataset.}
	\label{table:sampset_distrib}
	{\begin{tabular}{|>{\centering}m{2.2cm}|>{\centering}m{1.1cm}|>{\centering}m{1.1cm}|>{\centering}m{1cm}|p{1.3cm}|}
			\hline
			Dataset & Normal Samples & Anomaly Samples & Total Count & Anomaly\% \\
			\hline
			NSL-KDD SampleSet & 13460 & 1500 & 14960 & 10.03 \\
			\hline
			CIC-IDS2018 SampleSet & 20000 & 2230 & 22230 & 10.03 \\
			\hline
			TON\_IoT(Win10) SampleSet & 1948 & 217 & 2165 & 10.02 \\
			\hline			
			%		\caption{Sample Distribution for NSL-KDD Dataset}
	\end{tabular}}
	\vspace{-1em}
\end{table}

For the proposed framework, we perform an independent experiment for each SampleSet and following is a description of the experimental process:
We apply 3-fold CV which provides us with the training and test sets for each SampleSet. The training phase in our proposed framework involves the first two stages as detailed in Section III. In Stage 1, the training data is assumed as unlabeled and our clustering methodology is applied to first identify the normal samples. The output labeled normal samples is then used as training input to the GANomaly detector in Stage 2.
In the testing phase, the above trained GANomaly detector will be directly used to label the data in the test sets.

For the third stage of the framework, each anomaly will be further classified into its corresponding attack category and therefore, individual attack categories are considered for the third stage. The details regarding the training and evaluation of the CNN in the third stage are discussed in the Sec.\ref{sec:cnn_res}.

\subsection{Evaluation Metrics}
For the anomaly detection, there are multiple evaluation metrics available such as accuracy, precision, recall etc. As the number of anomalous samples is smaller compared to the normal samples, the accuracy may not be an ideal metric. We adopt True Positive Rate (TPR), False Positive Rate (FPR), and area under the Receiver Operating Characteristics curve (AUC) as evaluation metrics.

True positive rate is the ratio of the number of data samples correctly predicted as attacks to the total number of attack samples present in the dataset. True positives refer to the number of attack samples correctly predicted as attacks by the model and false negatives are the number of the attack samples incorrectly predicted as normal by the given model. False positive rate is the ratio of the number of normal samples incorrectly identified as attacks to the total number of normal samples present in the dataset. False positives are the number of the normal samples incorrectly predicted as attacks by the model, and true negatives are the number of the normal samples correctly predicted as normal by the model. In addition, receiver operating characteristic(ROC) curve~\cite{roc} is used to evaluate the anomaly detection results.

For evaluating the CNN, we will calculate the log-loss for the multi-class classification.

\subsection{Baselines}
The baseline for the performance will be established using the KMeans algorithm, the OneClass Support Vector Machine (OCSVM) algorithm and the GANomaly algorithm. We will now briefly explain baseline algorithms.\\
\noindent \textbf{KMeans Clustering:} KMeans clustering method in general is described in Sec.\ref{sec:kmeans}. For baseline only KMeans clustering, first the KMeans clustering is performed on the input data. Based on the predefined threshold of cluster size, clusters with size smaller than the threshold are identified as anomaly clusters and all the samples within that cluster are identified as anomalies \cite{casas2012unsupervised}.
By setting different cluster size threshold, distinct TPR and FPR values are obtained to plot ROC curve.\\
\noindent \textbf{OCSVM:} OCSVM method is an extension of Support Vector Machine (SVM) method. It is a method suitable for handling unlabeled data \cite{ocsvm}.\\
\noindent \textbf{GANomaly:} The GANomaly network used for baseline comparison is identical to the one described in Sec.\ref{sec:gano} in the proposed framework.

For baseline OCSVM and GANomaly, the entire training set will be used during the training phase. During the test phase, the test dataset will be used for the evaluation.

\subsection{System Setup}
The entire framework is implemented in Python. Even though the Pytorch source code for GANomaly is available in \cite{akcay2018ganomaly}, that implementation is only suitable for 2D image data. Therefore, we have implemented the GANomaly suitable for 1D data in Keras with Tensorflow backend. Similarly, CNN in the third stage is implemented in Keras with Tensorflow backend.
Experiments were performed on the system with Ubuntu 20.04 OS, 16-core CPU, 64GB RAM and Nvidia RTX 3070 GPU.
The time taken for anomaly detection per sample is about 130-140ms. For CNN, the resampling notably increases the training time as the number of training samples increases significantly. The training time per epoch for NSL-KDD dataset, CIC-IDS2018 and TON\_IoT dataset is about 32s, 35s and 8s, respectively.

\section{Results}
We will first evaluate the anomaly detection with the output from the stage 2 of the framework through the AUC and TPR-FPR comparison.
\subsection{Anomaly Detection}

The mean of the results by the area under the ROC curve (AUC) for each method and for all the three datasets across the different SampleSets  is summarized in Table \ref{table:auroc}. The bold values indicate the maximum AUC value for that dataset.
%\vspace{-0.7em}
\begin{table}[htbp]
	\centering
	\caption{Area under the ROC curve (AUC)}
	\label{table:auroc}
	%	\resizebox{\columnwidth}{!}{\begin{tabular}{|c|c|c|c|c|}
			{\begin{tabular}{|c|>{\centering}m{1.1cm}|>{\centering}m{1.3cm}|>{\centering}m{1.1cm}|p{1.1cm}|}
					\hline
					%		\diagbox{\rotatebox{10}{\textbf{Dataset}}}{\rotatebox{10}{\textbf{Method}}} & \textbf{KMeans} & \textbf{GANomaly} & \textbf{OCSVM} & \textbf{Proposed}\\
					\diagbox{\textbf{Dataset}}{\textbf{Method}} & \textbf{KMeans} & \textbf{GANomaly} &\textbf{OCSVM} & \textbf{Proposed Method}\\
					\hline
					NSL-KDD & 0.794 & 0.887 & \textbf{0.977} & 0.916\\ 
					\hline
					CIC-IDS2018 & 0.661 & 0.688 & 0.446 & \textbf{0.703} \\ 
					\hline
					TON\_IoT(Win10) & 0.793 & 0.889 & 0.756 & \textbf{0.918} \\ 
					\hline
					%		\caption{Sample Distribution for NSL-KDD Dataset}
			\end{tabular}}
			\vspace{-1em}
		\end{table}
		
		It can be observed from the results, that the AUC of the proposed framework is generally better than rest of the algorithms, except for the case of OCSVM algorithm in the NSL-KDD dataset. The relatively inferior performance of OCSVM in case of CIC-IDS2018 and TON\_IoT dataset could be attributed to the presence of anomalous samples in the training data. Ideally, the OCSVM algorithm should trained using samples from only a single class for effective anomaly detection. Since that is not the case for our training set, the performance of OCSVM is relatively inferior than our proposed algorithm. Interestingly, despite the presence of anomalous samples in the training set, the OCSVM still outperforms rest of the algorithms in case of NSL-KDD dataset. It is possible that specifically in case of NSL-KDD dataset, the number of normal samples in the training set are sufficient for OCSVM to perform effective anomaly detection.
		
		Even though the baseline GANomaly is same as the GANomaly used in the Stage 2 of the proposed framework, the performance of the proposed framework is consistently better than the baseline GANomaly. As mentioned in Sec. \ref{sec:gano}, the training of GANomaly needs to be performed using the non-anomalous samples. The GANomaly in our framework is trained with the normal samples identified from the training set in the Stage 1 of our framework whereas the baseline GANomaly is trained using the entire training dataset, which explains the better performance of our proposed framework. 
		
		The comparison results for the overall mean TPR-FPR are shown in Table \ref{table:tpr}. The bold values within a row indicate the highest TPR and lowest FPR for that particular dataset. KMeans and OCSVM both have highest TPR for NSL-KDD dataset. Though KMeans algorithm has the highest TPR even for CIC-IDS2018 and TON\_IoT(Windows 10) datasets, its FPR is also high for those datasets. GANomaly provides lowest FPR for CIC-IDS2018 dataset. Proposed method has the lowest FPR for NSL-KDD and TON\_IoT datasets.
		%Normally higher TPR results in higher FPR and lower TPR results in lower FPR, which is validated in Table \ref{table:tpr}.
		Our proposed method generally performs comparably or relatively better than the state-of-the-art methods.
		
		\begin{table*}[htb]
			\centering
			\begin{minipage}{0.65\textwidth}
				\centering
				\caption{TPR-FPR Comparison}
				\label{table:tpr}
				%		\resizebox{\columnwidth}{!}{\begin{tabular}{|c|c|c|c|c|c|c|c|c|}
						\begin{tabular}{|c|c|c|c|c|c|c|c|c|}%
							\cline{2-9}
							\multicolumn{1}{c|}{} & \multicolumn{8}{|c|}{Method} \\
							\cline{2-9}
							\multicolumn{1}{c|}{} & \multicolumn{2}{|c|}{KMeans} & \multicolumn{2}{|c|}{GANomaly} & \multicolumn{2}{|c|}{OCSVM} & \multicolumn{2}{|c|}{Proposed Method} \\%
							\cline{1-9}
							\textbf{Dataset} & TPR & FPR & TPR & FPR & TPR & FPR & TPR & FPR \\ %
							%					\specialrule{1pt}{1pt}{1pt}
							\cline{1-9}
							NSL-KDD & \textbf{0.999} & 0.367 & 0.859 & 0.163 & \textbf{0.999} & 0.444 & 0.865 & \textbf{0.132} \\ 
							\hline
							CIC-IDS2018 & \textbf{0.717} & 0.368 & 0.618 & \textbf{0.219} & 0.472 & 0.504 & 0.677 & 0.243 \\  
							\hline
							TON\_IoT(Win10) & \textbf{0.998} & 0.34 & 0.86 & 0.19 & 0.977 & 0.448 & 0.872 & \textbf{0.142} \\
							\hline					
							%		\caption{Sample Distribution for NSL-KDD Dataset}
						\end{tabular}
						%		\setlength{\belowcaptionskip}{-8pt}
						%		\caption{TPR-FPR Comparison}
						%		\label{table:tpr}
					\end{minipage}
					\begin{minipage}{0.3\textwidth}
						\centering
						\caption{Log-loss for Stage 3: CNN}
						\label{table:cnn_log}
						\begin{tabular}{|c|m{1cm}|m{1cm}|}
							\hline
							Dataset & Base Log-loss & CNN Log-loss \\
							\hline
							NSL-KDD & 0.5931 & 0.1787 \\
							\hline
							CIC-IDS2018 & 1.5268 & 0.1388 \\
							\hline
							TON\_IoT (Win10) & 1.375 & 0.013 \\
							\hline
							
							%		\caption{Sample Distribution for NSL-KDD Dataset}
						\end{tabular}
					\end{minipage}
					\vspace{-2em}
				\end{table*}

\subsection{CNN Results} \label{sec:cnn_res}
The training process of the CNN in the third stage is performed independently. The number of samples for each dataset prior to resampling is shown in Table \ref{tab:cnn_samp}.
%The anomalous samples used for training of the CNN are not part of anomalous samples in SampleSets used for the evaluation of the previous stages.
For NSL-KDD, before resampling, the number of training samples for U2R is significantly lower than that of DoS attacks.
\vspace{-0.5em}
\begin{table}[H]
	\centering
	\caption{Number of Samples for each dataset. From left to right: NSL-KDD, CIC-IDS2018, TON\_IoT}
	\label{tab:cnn_samp}
	\begin{tabular}{|c|c|}
		\hline
		\textbf{Type} & \textbf{Samples} \\
		\hline
		DoS & 44371 \\ 
		\hline
		Probe & 11356 \\ 
		\hline
		R2L & 925 \\ 
		\hline
		U2R & 32 \\ 
		\hline
		%		\caption{Sample Distribution for NSL-KDD Dataset}
	\end{tabular}
	\quad
	\begin{tabular}{|c|c|}
		\hline
		\textbf{Type} & \textbf{Samples} \\
		\hline
		Infiltration & 17000 \\ 
		\hline
		DoS & 58000 \\ 
		\hline
		Bruteforce & 34000 \\  
		\hline
		Web & 784 \\
		\hline
		Bot & 17000 \\ 
		\hline
		DDoS+PortScan & 35300 \\ 
		\hline
		%			\caption{Sample Distribution for NSL-KDD Dataset}
	\end{tabular}
	\quad
	\begin{tabular}{|c|c|}
		\hline
		\textbf{Type} & \textbf{Samples} \\
		\hline
		DDoS & 3775 \\ 
		\hline
		DoS & 375 \\ 
		\hline
		Injection & 475 \\ 
		\hline
		XSS & 750 \\ 
		\hline
		Password & 2625 \\ 
		\hline
		Scanning & 325 \\ 
		\hline
		MITM & 57 \\ 
		\hline
	\end{tabular}
	%\caption{Number of Samples for each dataset. From left to right: NSL-KDD, CIC-IDS2018, TON\_IoT}
	%\label{tab:cnn_samp}
	%\vspace{-1.2em}
\end{table}

For CIC-IDS2018 dataset, the 14 attack types are classified into 6 broad attack categories as per the dataset details\cite{sharafaldin2018toward}. The number of training samples available for Web Attack is significantly lower compared to the other attack types.

For TON\_IoT Windows10, before resampling, very few samples for MITM attack are available for training compared to rest of the attack types.	
%\begin{table}[htbp]
%	\centering
%	\caption{Sample Distribution for TON\_IoT (Windows 10) Dataset}
%	\label{table:ton_samp}

As mentioned in Sec. \ref{sec:cnn}, for such cases of imbalanced datasets, resampling technique such as ADASYN are useful to improve the sample availability for minority classes during the training. After the resampling technique ADASYN is applied, the number of samples for each attack category was approximately made equal to the maximum number of samples for a single category prior to the resampling.

The log-loss results for CNN are shown in Table \ref{table:cnn_log}. To show the classification effectiveness of the CNN model, we are comparing the CNN log-loss against the na\"ive or base log-loss. The base log-loss is calculated by making the predictions based on the sample distributions in the training set for each dataset, prior to the resampling. The base log-loss value depends on the number of classes and the prevalence of the samples in each class in training dataset.

As seen in the Table \ref{table:cnn_log}, the CNN log-loss is better than the base log-loss for all three datasets. The lower NN log-loss values compared to the base log-loss indicate that the prediction probabilities of CNN are closer to the corresponding true values. To examine the CNN classification in more detail, we will now look at the confusion matrix for each test dataset.

\begin{figure}[htb]
	\centering
	\includegraphics[trim=2cm 0cm 2cm 0cm, width=1.3in]{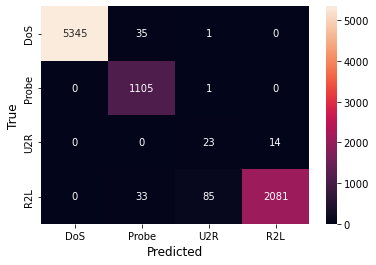}
	\setlength{\belowcaptionskip}{-8pt}
	\caption{Confusion Matrix for CNN: NSL-KDD Dataset}
	\label{fig:confm_nsl}
%		\vspace{-1em}
\end{figure}

\begin{figure}[htb]
	\centering
	\includegraphics[trim=0cm 0cm 0cm 0cm, width=2.6in]{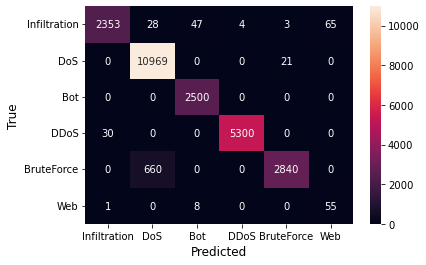}
	\setlength{\belowcaptionskip}{-8pt}
	\caption{Confusion Matrix for CNN: CIC-IDS2018 Dataset}
	\label{fig:confm_ids}
%	\vspace{-1em}
\end{figure}

\begin{figure}[htb]
	\centering
	\includegraphics[trim=0cm 0cm 0cm 1cm, width=2.5in]{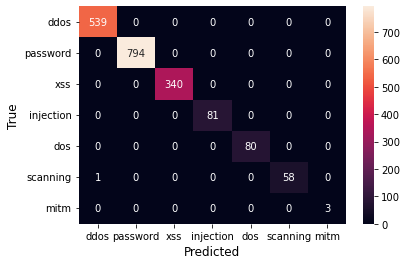}
	\setlength{\belowcaptionskip}{-8pt}
	\caption{Confusion Matrix for CNN: TON\_IoT (Windows10) Dataset}
	\label{fig:confm_ton}
\end{figure}

The attack category classification for each test dataset is shown in Fig. \ref{fig:confm_nsl}, \ref{fig:confm_ids} and \ref{fig:confm_ton} for NSL-KDD, CIC-IDS2018 and TON\_IoT (Win10) datasets, respectively. The true labels are on the Y-axis and the predicted labels are on the X-axis. The test dataset for the evaluation of CNN stage is different from the test dataset of the anomaly detection stage.
%There is no overlap between Anomaly samples from SampleSets and CNN training dataset. However, there may be an overlap between Anomaly test dataset and the CNN evaluation test dataset if the number of samples available for that attack category is low. 

For NSL-KDD, the test dataset within the NSL-KDD is used as the test dataset for the NSL-KDD evaluation. As shown in Fig. \ref{fig:confm_nsl}, despite having low number of samples prior to resampling in the training set, R2L classification shows decent accuracy. The classification for the U2R is least accurate ($\approx$62\%) which can be a result of its extremely low number of samples available in the training set prior to oversampling.

For CIC-IDS2018, the classification across different classes is generally accurate as shown in Fig \ref{fig:confm_ids}.
%After generating the training set from the CIC-IDS2018, the test set is generated randomly from the remaining samples available in CIC-IDS2018 and its size is about 15\% of the training set.
The only significant mis-classifcation is for the Bruteforce attack samples getting classified as DoS attacks. As both attack categories had sufficient samples in the training dataset prior to oversampling, the reason for the mis-classification could be attributed to the similarity in their features.

For TON\_IoT (Windows10), the number of available samples are split as 70\%-30\% for training and test dataset prior to the resampling process. As shown in Fig.\ref{fig:confm_ton}, the test performance achieves $\approx$99\% accuracy as the CNN is able to efficiently capture the features within the dataset.
%\vspace{-0.5em}
\section{Conclusion}
We proposed a three-stage deep learning anomaly based intrusion detection framework by integrating unsupervised (K-means clustering), semi-supervised (GANomaly) and supervised (CNN) learning methods. 
We implemented and evaluated the performance of our framework on three datasets: NSLKDD, CIC-IDS2018, and TON IOT (Win10). Our proposed work achieved better FPR with comparable TPR performance when evaluated against state-of-the-art methods. As future work, parallelization and utilization of advanced machine learning algorithms can be explored for further performance enhancement.

%Be sure that the 
%symbols in your equation have been defined before or immediately following 
%the equation. Use ``\eqref{eq}'', not ``Eq.~\eqref{eq}'' or ``equation \eqref{eq}'', except at 
%the beginning of a sentence: ``Equation \eqref{eq} is . . .''

%\section*{References}
%
%Please number citations consecutively within brackets \cite{b1}. The 
%sentence punctuation follows the bracket \cite{b2}. Refer simply to the reference 
%number, as in \cite{b3}---do not use ``Ref. \cite{b3}'' or ``reference \cite{b3}'' except at 
%the beginning of a sentence: ``Reference \cite{b3} was the first $\ldots$''
%
%Number footnotes separately in superscripts. Place the actual footnote at 
%the bottom of the column in which it was cited. Do not put footnotes in the 
%abstract or reference list. Use letters for table footnotes.
%
%Unless there are six authors or more give all authors' names; do not use 
%``et al.''. Papers that have not been published, even if they have been 
%submitted for publication, should be cited as ``unpublished'' \cite{b4}. Papers 
%that have been accepted for publication should be cited as ``in press'' \cite{b5}. 
%Capitalize only the first word in a paper title, except for proper nouns and 
%element symbols.

%\vspace{12pt}
%\color{red}
%IEEE conference templates contain guidance text for composing and formatting conference papers. Please ensure that all template text is removed from your conference paper prior to submission to the conference. Failure to remove the template text from your paper may result in your paper not being published.

\end{document}